\documentclass[twocolumn,aps,showpacs,prb,tightenlines,amsmath,amssymb]{revtex4}
\usepackage{bm}
\usepackage{graphicx}              
\usepackage{amssymb}               
\usepackage{amsmath}                     
\usepackage{colordvi}
\usepackage{calrsfs} \newcommand{\bgreek}[1]{\mbox{\boldmath$#1$\unboldmath}}
\begin{document}   

\title{Spin relaxation in ultracold spin-orbit coupled $^{40}$K gas}
\author{T. Yu}
\author{M. W. Wu}
\thanks{Author to whom correspondence should be addressed}
\email{mwwu@ustc.edu.cn.}
\affiliation{Hefei National Laboratory for Physical Sciences at
  Microscale and Department of Physics, 
University of Science and Technology of China, Hefei,
  Anhui, 230026, China} 
\date{\today}

\begin{abstract} 
We report the anomalous D'yakonov-Perel' spin relaxation in ultracold spin-orbit
coupled $^{40}$K gas when the coupling between $|9/2,9/2\rangle$ and
$|9/2,7/2\rangle$ states is much stronger than the
spin-orbit coupled field. Both the transverse and longitudinal spin relaxations are investigated with small and
large spin polarizations. It is found that with small spin polarization, the
transverse (longitudinal) spin relaxation is divided into four (two) regimes: the normal weak
scattering regime, the anomalous D'yakonov-Perel'-like regime,
the anomalous Elliott-Yafet-like regime and the normal strong scattering regime (the anomalous Elliott-Yafet-like regime and the normal strong scattering
regime). With large spin polarization, we find that the Hartree-Fock self-energy, which acts as an
effective magnetic field, can markedly suppress the transverse spin
relaxation in both weak and strong scattering limits. Moreover, by noting that
as both the momentum relaxation time and the Hartree-Fock effective magnetic field vary with the
scattering length in cold atoms, the anomalous D'yakonov-Perel'-like regime is suppressed and the transverse spin relaxation
is hence divided into three regimes in the scattering length
dependence: the normal weak scattering regime, the anomalous
Elliott-Yafet-like regime and the strong scattering regime. On the other hand,
the longitudinal spin relaxation is again divided into the anomalous Elliott-Yafet-like and normal strong scattering
regimes. Furthermore, it is
found that  the
longitudinal spin relaxation can be either enhanced or suppressed by the Hartree-Fock effective magnetic field if the spin polarization is parallel or antiparallel to the
effective Zeeman magnetic field.  

\end{abstract}

\maketitle 

\section{Introduction} 
Recently, the synthetic gauge fields in Bose and Fermi cold atom systems have been experimentally realized by laser control technique, which opens the door to study the spin-orbit
coupling (SOC) and related phenomenon in these systems.\cite{gauge1,gauge2,gauge3,gauge4,Bose,K,Li} Much attention has been attractted to the spin-orbit coupled Bose-Einstein condensate\cite{Bose}
and Fermi gas\cite{K,Li} after the pioneering experimental works with precisely controlled SOC. In the spin-orbit
coupled Bose-Einstein condensate, the SOC provides new possibilities to give rise to new
quantum phases.\cite{Wang2,Wang3,Wang4,PRB84,PRL108,PRA85,PRA86,PRL110,Rb} In the ultracold Fermi gas system,
the experimental realization of the spin-orbit coupled $^{40}$K (Ref.~\onlinecite{K}) and $^6$Li (Ref.~\onlinecite{Li})
systems provide a platform to simulate the spin dynamics in the solid state
systems. Moreover, with tunable scattering strength between atoms by the Feshbach
resonance,\cite{Feshbach} the spin-orbit
coupled ultracold Fermi gas can be used to study the
influence of the interparticle interaction on the spin relaxation without being disturbed by disorders, inevitable in solid state systems.\cite{sherman_cold,Dassarma} 

The study of the spin-dynamics in the spin-orbit coupled Fermion gas has just
started. Experimentally, in the $^{40}$K system, the lowest two magnetic sublevels $|9/2,9/2\rangle$ and
$|9/2,7/2\rangle$, which are referred to as two spin-1/2 states, are coupled by a pair of Raman
beams with the coupling strength $\Omega$.\cite{K} In this system, the
effective Hamiltonian including the SOC can be written as ($\hbar \equiv 1$ throughout this paper)\cite{Bose,K,Li,Hamiltonian}
\begin{equation}
\hat{H}_0=\varepsilon_{\bf k}+{\bf{\Omega(k)}}\cdot\bgreek{\sigma}/2,
\label{Hamiltonian}
\end{equation}
with the effective magnetic field
\begin{equation}
{\bf{\Omega(k)}}=\Big({\Omega},0,-{\delta}-{2k_r k_x}/{m}\Big).
\label{field}
\end{equation}
The first term $\varepsilon_{\bf k}={k^2}/(2m)$ in
 Eq.~(\ref{Hamiltonian}) is the kinetic energy of atom
with ${\bf k}$ representing the center-of-mass momentum and $m$ standing for the mass of the
atom. $\Omega_z({\bf k})$ in the second term is the effective magnetic field created by the Raman beams
with $k_r$ being the recoil momentum of the Raman beam and $\delta$ denoting the Raman
detuning. $\bgreek \sigma$ is the vector composed
of the Pauli matrices. It is noted that the coupling strength $\Omega$ acts as an effective Zeeman
magnetic field along the $\hat{\bf x}$-direction and the $\bf k$-dependent effective
magnetic field created by the Raman beams in Eq.~(\ref{field}) is perpendicular to the effective Zeeman
magnetic field. Furthermore, $\delta$ can be set to be zero, 
which can be realized by
tuning the Raman beams.\cite{Bose,K,Li}
 It is interesting to see that the Hamiltonian
Eq.~(\ref{Hamiltonian}) with $\delta=0$ is similar to that of (110) semiconductor quantum wells with external
magnetic field parallel to the $\hat{x}$-axis.\cite{Zhou} 

Apart from the experimental investigations, the spin dynamics has been studied partly based on the effective Hamiltonian [Eq.~(\ref{Hamiltonian})] theoretically.  Tokatly and Sherman suggested that the
polarized spin states are created by a weak effective Zeeman magnetic field, and the
polarized spins relax to the equilibrium state due to the SOC and interatom collisions after swiching off the Zeeman
magnetic field.\cite{sherman_cold} Natu and Das Sarma explored the spin dynamics of the two-dimensional,
non-degenerate, harmonically trapped ultracold Fermi gas in the non-interaction
and diffusive limits, in the absence of the effective Zeeman magnetic field.\cite{Dassarma} They reported oscillations of the spin polarization in time domain and spin helix in space domain. So far, the spin dynamics with strong effective Zeeman magnetic field in the experimental feasibility has not been studied in the Fermi ultracold atom system. 

The physics of the spin dynamics in the spin-orbit coupled $^{40}$K gas in the
presence of a strong effective Zeeman magnetic field can be very rich and
intriguing. This can be conjectured from a very recent study by Zhou {\it et
  al.} on the spin dynamics in (110) quantum wells in the presence of a strong
magnetic field ${\bf B}$ parallel to the $\hat{x}$-axis.\cite{Zhou} There, due to the unique form of the effective magnetic field ${\bf \Omega(k)}$, the spin relaxation may show anomalous scalings between the spin relaxation time (SRT) $\tau_{\rm s}$ and momentum relaxation time $\tau_p^{\ast}$ for the transverse (longitudinal) spin
relaxation with spins polarization perpendicular (parallel) to ${\bf B}$, which
is very different from the conventional situation with zero or small magnetic field.

In the conventional spin relaxation in
  n-type semiconductors, the relevant spin relaxation mechanisms are the
  Elliott-Yafet (EY)\cite{Yafet,Elliot} and D'yakonov-Perel' (DP) mechanisms\cite{DP}. In
the EY mechanism, due to spin mixing, electron spins have a small chance to flip
during each scattering, with the spin relaxation time (SRT) $\tau_{\rm s}$
proportional to the momentum relaxation time $\tau_p^{\ast}$. In the DP
mechanism, the Dresselhaus\cite{dres} and/or Rashba\cite{Rashbaterm} spin-orbit coupling (SOC) provide
a ${\bf k}$-dependent effective magnetic field ${\bf \Omega(k)}$ and electron spins decay due to
their precessions around this ${\bf k}$-dependent effective magnetic field during
the free flight between adjacent scattering events. Specifically, in the strong (weak) scattering limit when $\langle|\Omega({\bf
  k})|\rangle\tau_p^\ast\ll 1$ ($\langle
|\Omega({\bf k})|\rangle\tau_p^\ast\gtrsim 1$), 
$\tau_s$ is inversely
proportional (proportional) to $\tau_p^{\ast}$.

Interestingly, in the work of Zhou {\it et al.}, by varying the impurity density, the transverse (longitudinal) spin
relaxation can be divided into four (two) regimes: the normal weak
scattering regime ($\tau_{\rm s}\propto \tau_p^{\ast}$), the anomalous DP-like regime ($\tau_{\rm s}^{-1}\propto \tau_p^{\ast}$),
the anomalous EY-like regime ($\tau_{\rm s}\propto \tau_p^{\ast}$)
and the normal strong scattering regime ($\tau_{\rm s}^{-1}\propto \tau_p^{\ast}$) (the anomalous EY-like regime and the normal strong scattering regime).\cite{Zhou}  However, in the three-dimensional ultracold atom system, in spite of the similarity of the Hamiltonian, the interatom scattering, whose strength can also be tuned by the Feshbach resonance,\cite{Feshbach} is different from the impurity scattering. Whether the anomalous scalings in semiconductor systems are still valid is yet to be checked.

In the present work, with the interatom interaction explicitly included, we investigate the spin relaxation in the three-dimensional ultracold
$^{40}$K gas under the strong effective Zeeman magnetic field by the kinetic spin
Bloch equation (KSBE) approach both analytically and numerically.\cite{wu-review} Interestingly, similar anomalous scalings between the SRT and momentum relaxation time are observed in the spin relaxation under strong effective Zeeman magnetic field satisfying $|\Omega|\gg \langle|\Omega_z({\bf k})|\rangle$ when the spin polarization is small by tuning the interatom scattering strength. $\langle...\rangle$ here denotes the ensemble average. It is further revealed in this study that when the spin polarization is large, these anomalous relations are significantly influenced by the Hartree-Fock (HF) self-energy originating from the interatom interaction, acting as an effective magnetic field.\cite{jianhua23,jianhua36,jianhua37,jianhua46,wu-review} It is shown that due to this HF effective magnetic field, the transverse spin relaxation can be suppressed. Consequently, in the scattering length
dependence, by noting that both the momentum relaxation time and the HF effective magnetic field vary with the
scattering length in cold atoms, the anomalous DP-like regime is suppressed and the transverse spin relaxation
is hence divided into three instead of four regimes: the normal weak scattering regime, the anomalous
EY-like regime and the strong scattering regime. On the other hand, the longitudinal spin relaxation can be either enhanced or suppressed by the HF effective magnetic field when the spin polarization is parallel or antiparallel to the effective Zeeman magnetic field and its scattering length dependence is again divided into the anomalous EY-like and normal strong scattering
regimes. In addition, we find that the specific form of the effective magnetic field ${\bf \Omega(k)}$ leads to a strong anisotropy between the transverse and longitudinal spin relaxations.
 
This paper is organized as follows. In Sec.~II, we present the
model and KSBEs. The main results are given in
Sec.~III. In Sec.~III A,
 we solve the KSBEs analytically in a simplified model with the inelastic interatom scattering replaced by the elastic one and reveal the physics of the anomalous spin relaxation. In Sec.~III B, with the inclusion of the interatom scattering, we numerically study the anomalous spin relaxation by varying the interatom scattering strength. The anisotropy of the transverse and longitudinal spin relaxations is also addressed in this part. We
summarize in Sec.~IV.
\section{Hamiltonian and KSBEs}
We start the investigation from the full Hamiltonian of the spin-orbit coupled ultracold atom
systems, which is composed of the effective Hamiltonian $\hat{H}_0$ [Eq.~(\ref{Hamiltonian})] and
interatom interaction $\hat{H}_{\rm int}$,\cite{interaction,interaction2,Dassarma,pwave1,Spielman} 
\begin{equation}
\hat{H}=\hat{H}_0+\hat{H}_{\rm int}.
\end{equation}
The interaction Hamiltonian $\hat{H}_{\rm int}$ can be written
as\cite{interaction,interaction2,Dassarma,pwave1,Spielman}

\begin{eqnarray}
\nonumber
H_{\rm int}&=&\frac{1}{2V}\sum_{\sigma=\uparrow,\downarrow}\sum_{{\bf
    k_1+k_2=k_3+k_4}}g_{\sigma,\sigma}\Psi^{\dagger}_{{\bf
    k_4},\sigma}\Psi^{\dagger}_{{\bf k_3},\sigma}\Psi_{{\bf
    k_2},\sigma}\Psi_{{\bf k_1},\sigma}\\
&& {}+\frac{g_{\uparrow,\downarrow}}{V}\sum_{{\bf
    k_1+k_2=k_3+k_4}}\Psi^{\dagger}_{{\bf
    k_4},\uparrow}\Psi^{\dagger}_{{\bf k_3},\downarrow}\Psi_{{\bf k_2},\downarrow}\Psi_{{\bf k_1},\uparrow}.
\label{interaction}
\end{eqnarray}
Here $\Psi_{{\bf k},\sigma}$ is the annihilation operator of the atom with momentum
$\bf k$ in the spin-$\sigma$ state with $\sigma\equiv\uparrow,\downarrow$, 
$V$ is the volume of the system, and $g_{\uparrow,\downarrow}={4\pi a}/{m}$ is the
scattering potential with $a$ being the scattering length. The scattering potential $g_{\uparrow,\uparrow}$ and $g_{\downarrow,\downarrow}$ are set to be $g$.\cite{pwave1,Spielman} 

The KSBEs, derived via the nonequilibrium Green function
method with the generalized Kadanoff-Baym Ansatz,\cite{wu-review,2001,jianhua23,jianhua15,jianhua52} are utilized to study the spin relaxation in the ultracold Fermi gas:
\begin{equation}
  \partial_t \rho_{\bf k}(t)=\partial_t\rho_{\bf k}(t)|_{\rm coh}+\partial_t\rho_{\bf k}(t)|_{\rm  scat}.
\label{ksbe}
\end{equation}
In these equations, $\rho_{\bf k}(t)$  represent the density matrices of
atom with momentum ${\bf k}$  at
 time $t$, in which the diagonal elements $\rho_{{\bf k},\sigma\sigma}$ describe the atom distribution
 functions and the off-diagonal elements $\rho_{{\bf k},\sigma-\sigma}$ 
represent the spin coherence for the inter spin-band correlation. In the
  collinear space, the coherent term is given by 
\begin{equation}
\partial_t\rho_{\bf k}(t)|_{\rm
   coh}=-i\big[{\bf \Omega}({\bf
     k})\cdot{\bgreek\sigma}/2+\Sigma_{\rm HF},\rho_{\bf
   k}(t)\big],
\end{equation}
 where $\Sigma_{\rm HF}=-g\sum_{\bf k'}\rho_{\bf k'}$ is the HF self-energy and $[\ ,\ ]$ denotes the commutator. The scattering terms
 $\partial_t\rho_{\bf k}(t)|_{\rm  scat}$ represent
 the interatom scattering. In our study, we focus on the
situation that the effective Zeeman splitting energy and the 
SOC energy are much  smaller than the Fermi energy 
and hence the scattering terms read\cite{Jianhua}
\begin{eqnarray}
\nonumber
&&\partial_t\rho_{\bf k}(t)|_{\rm  scat}\\
\nonumber
&&\mbox{}=-\pi g^2\sum\limits_{{\bf k'},{\bf k''}}\delta(\varepsilon_{\bf
k'}-\varepsilon_{\bf k}+\varepsilon_{\bf k''}-\varepsilon_{\bf k''-k+k'})\\
\nonumber
&&\mbox{}\times \big[\rho_{\bf k'}^{>}\rho_{\bf k}^{<}\mbox{Tr}(\rho_{\bf
  k''-k+k'}^{<}\rho_{\bf k''}^{>})-\rho_{\bf k'}^{<}\rho_{\bf k}^{>}\mbox{Tr}(\rho_{\bf
  k''-k+k'}^{>}\rho_{\bf k''}^{<})\big]\\
&&\mbox{}+{\rm H.c.},
\end{eqnarray}
with $\rho_{\bf k}^{<}=\rho_{\bf k}$ and $\rho_{\bf k}^{>}=1-\rho_{\bf k}$.

By solving the KSBEs, one
obtains the SRT from the time evolution of the spin polarization
$P(t)=\sum_{\bf k}\mbox{Tr}[\rho_{\bf
    k}(t) {\bgreek \sigma}\cdot \hat{\bf n}]/n_a$ with $\hat{\bf n}$ standing for the spin
  polarization direction and $n_a$ being the density of the ultracold atom. 
The initial condition is set to be
\begin{equation}
\rho_{\bf k}(0)=\frac{F_{{\bf k}\uparrow}+F_{{\bf k}\downarrow}}{2}+\frac{F_{{\bf k}\uparrow}-F_{{\bf k}\downarrow}}{2}{\bgreek \sigma}\cdot \hat{\bf n},
\label{polarization}
\end{equation}
where $F_{{\bf k}\sigma}=\{\exp[(\varepsilon_{{\bf k}}-\mu_{\sigma})/(k_BT)]+1\}^{-1}$ is the Fermi distribution function at temperature $T$, with $\mu_{\uparrow,\downarrow}$ standing for the chemical potentials determined by the atom
density $n_a$=$\sum_{\bf k}$Tr[${\rho_{\bf k}}$] and the initial
spin polarization $P(0)$.
\section{Results}
\normalsize
In this section, we solve the KSBEs first analytically and then fully numerically. In order to reveal the physics of the anomalous spin relaxation under strong effective Zeeman magnetic field, we first investigate the spin relaxation in a simplified model analytically, with the inelastic interatom scattering replaced by the elastic one. The situations with both small and large spin polarizations are considered. Specifically, the effects of the HF effective magnetic field on the transverse and longitudinal spin relaxations are shown explicitly. Then we numerically study the anomalous spin relaxation with the genuine interatom scattering. The scattering strength dependence of the spin
relaxation is discussed facilitated with a full understanding of the SRTs from the analytical model.

\subsection{Analytical results}
\label{section1}
\begin{figure}[ht]
  {\includegraphics[width=8cm]{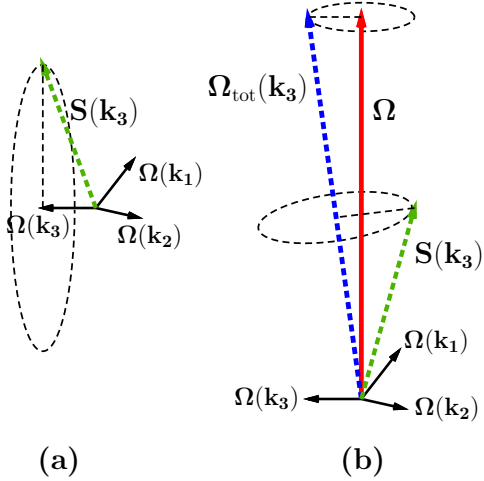}}
  \caption{(Color online) Schematic of the DP mechanism of the spin relaxation under the weak
    (a) and strong (b) effective Zeeman magnetic fields. Due to the random momentum scattering, the ${\bf k}$-dependent
    magnetic field changes from ${\bf \Omega(k_1)}$ to ${\bf \Omega(k_3)}$ and
    the spin vector ${\bf S(k)}$ precesses around the ${\bf k}$-dependent
    magnetic field during the free flight between adjacent
    scattering events. With the weak effective Zeeman magnetic field, this
    causes the 
    conventional DP spin relaxation. With the strong effective Zeeman magnetic
    field satisfying $\Omega\gg \langle |\Omega({\bf k})| \rangle$, ${\bf
      \Omega}_{\rm tot}({\bf k})$ is nearly parallel to the effective Zeeman
    magnetic field ${\bf \Omega}=\Omega \hat{\bf z}$ and the DP spin relaxation shows anomalous behaviors.}
\label{figyw1}
\end{figure}

Before performing the numerical study by solving the KSBEs, we first
investigate the spin relaxation in a simplified model by replacing the inelastic
interatom scattering by the elastic one. The conventional spin relaxation
determined by the DP mechanism under the weak effective Zeeman magnetic field
with $|\Omega|\ll \langle|\Omega_z({\bf k})|\rangle$ has been studied in
semiconductors in the strong and weak scattering limits $\langle|\Omega_z({\bf
  k})|\rangle\tau_p^{\ast}\ll 1$ and $\langle|\Omega_z({\bf
  k})|\rangle\tau_p^{\ast}\gg 1$,
respectively,\cite{DP,Awschalom,Zutic,fabian565,Dyakonov,Korn,wu-review}
[this situation is schematically shown in Fig.~\ref{figyw1}(a)]. Accordingly, in the strong (weak) scattering limit, the SRT is
inversely proportional (proportional) to the momentum relaxation time. Moreover,
the influence of the HF effective magnetic field to the spin relaxation has been
fully investigated.\cite{jianhua23,Jianhua,jianhua36,jianhua37,wu-review}
Specifically, the SRT can be effectively enhanced due to the suppression of the
inhomogeneous broadening\cite{2001} [i.e., the ${\bf k}$-dependent effective
magnetic field $\Omega_z{(\bf k)}$] by the HF effective magnetic
field.\cite{jianhua23,jianhua36,jianhua37} Here, we focus on the spin relaxation
under the strong effective Zeeman magnetic field satisfying $|\Omega|\gg
\langle|\Omega_z({\bf k})|\rangle$ [illustrated in Fig.~\ref{figyw1}(b)].

The KSBEs including the HF effective magnetic field can be written as\cite{Bloch}
\begin{eqnarray}
\nonumber
&&\partial_t \rho_{\bf k}+{i}[\frac{\Omega_z({\bf k})}{2}\sigma_z,\rho_{\bf k}]+i[\frac{\Omega}{2}\sigma_x,\rho_{\bf
  k}]+i[\frac{{\bf \Omega}_{\rm HF}}{2}\cdot{\bgreek \sigma},\rho_{\bf
  k}]\\
&&\mbox{}+\sum\limits_{\bf k'}W_{\bf kk'}(\rho_{\bf k}-\rho_{\bf k'})=0,
\label{HF_term}
\end{eqnarray}
in which $W_{\bf kk'}={2\pi}|U_{\bf k-k'}|^2\delta(\varepsilon_{\bf
  k}-\varepsilon_{\bf k'})$ describes the elastic scattering with $|U_{\bf k-k'}|$ being the
scattering matrix element. In our study, by considering the property of the
  interatom scattering potential [Eq.~(\ref{interaction})], $|U_{\bf k-k'}|$ is set to be
  constant $g$ and hence $W_{\bf kk'}$ is proportional to $g^2$. $\Omega_z({\bf k})=\alpha k_x$ with $\alpha\equiv -2k_r/m$
 is the ${\bf k}$-dependent magnetic field and  
\begin{equation}
{\bf \Omega}_{\rm HF}=-{2g}\sum_{{\bf k}}\frac{1}{2}\mbox{Tr}\big[\rho_{\bf k}{\bgreek \sigma}\big]\equiv -{2g}\sum_{{\bf k}} {\bf S_k}
\label{HF}
\end{equation}
acts as the HF effective magnetic field. It is noted that unlike that in semiconductors, \cite{jianhua23,wu-review}the HF effective magnetic field here is always proportional to the total spin vector ${\bf S}=\sum_{{\bf k}}{\bf S_k}$.

\subsubsection{SRT in the absence of HF effective magnetic field}
When the spin polarization is small, i.e., the HF effective magnetic field is negligible, the total spin vector for the transverse and longitudinal spin relaxations can be obtained by solving the KSBEs [Eq.~(\ref{HF_term})] with large effective Zeeman magnetic field $|\Omega|\gg \langle|\Omega_z({\bf k})|\rangle$ in the strong scattering limt $\langle|\overline{{\Omega^2_z({\bf k})}}|/{(2\Omega)}\big\rangle\tau_{k,2}\ll 1$ (Ref.~\onlinecite{Zhou}):\cite{two-dimensional}
\begin{equation}
{\bf S}^{T}(t)\approx|{\bf S}^{T}(0)|e^{-t/\tau_{sz}}\big[\cos(\Omega t)\hat{\bf z}-\sin(\Omega t)\hat{\bf y}\big],
\label{TT}
\end{equation}
and
\begin{equation}
{\bf S}^{L}(t)\approx|{\bf S}^{L}(0)|\hat{\bf x}e^{-t/\tau_{sx}}.
\label{LL}
\end{equation}
The corresponding transverse and longitudinal SRTs are
\begin{eqnarray}
\label{z_relaxation}
\tau_{{\rm s}z}^{-1}&=&\tau_{{\rm s}y}^{-1}=\left\langle\frac{\overline{{\Omega^2_z({\bf
    k})}}^2\tau_{k,2}}{4\Omega^2}+\frac{{\overline{\Omega_z({\bf
    k})}}^2\tau_{k,1}}{2(1+\Omega^2\tau_{k,1}^2)}\right\rangle,\\
\tau_{{\rm s}x}^{-1}&=&\left\langle\frac{{\overline{\Omega_z({\bf
    k})}}^2\tau_{k,1}}{1+\Omega^2\tau_{k,1}^2}\right\rangle,
\label{x_relaxation}
\end{eqnarray}
in which $\overline{A_{\bf k}} = A_{\bf k}-\frac{1}{4\pi}\int d \omega_{\bf
    k}\,A_{\bf k}$ with $\omega_{\bf
    k}$ being the solid angle for momentum ${\bf k}$ and the momentum
relaxation time. 
\begin{equation}
\tau_{k,l}^{-1}=\frac{m\sqrt{k}g^2}{2\pi}\int_0^{\pi}\sin\theta d\theta\big[1-P_l(\cos\theta)\big],
\label{momentum}
\end{equation}
with $P_l(\cos\theta)$ being the Legendre function. Obviously, the momentum
relaxation time is inversely proportional to $g^2$ and hence $a^2$, which can be tuned by the Feshbach
resonance.\cite{Feshbach} Accordingly, the anomalous behavior of the transverse and longitudinal spin relaxations can be revealed.\cite{Zhou}

For the transverse spin relaxation, in the strong scattering limit, the SRT is limited by the competition of the two terms $\langle\overline{{\Omega^2_z({\bf
    k})}}^2\tau_{k,2}/(4\Omega^2)\rangle$ and $\langle{\overline{\Omega_z({\bf
    k})}}^2\tau_{k,1}/\big[2(1+\Omega^2\tau_{k,1}^2)\big]\rangle$ in Eq.~(\ref{z_relaxation}). Specifically, when 
\begin{equation}
\langle\overline{{\Omega^2_z({\bf
    k})}}^2\rangle\tau_{k,1}\tau_{k,2}/\langle{2\overline{\Omega_z({\bf
    k})}}^2\rangle\gg 1,
\label{strong_limit}
\end{equation} 
the first term is dominant and the SRT $\tau_{{\rm s}z}\approx\langle\overline{{\Omega^2_z({\bf
    k})}}^2\tau_{k,2}/(4\Omega^2)\rangle^{-1}$ is inversely proportional to the momentum relaxation time. It is further noted that when Eq.~(\ref{strong_limit}) is satisfied, for the situation under the weak effective Zeeman magnetic field, the system is in the weak scattering limit [$\langle|\Omega_z({\bf
      k})|\rangle\tau_{k,1}\gg 1$] with the SRT proportional to the
  momentum relaxation time. It is on this sense, this behavior under the strong
  effective Zeeman magnetic field is anomalous and this regime is referred to as
  the anomalous DP-like regime.\cite{Zhou} When 
\begin{equation}
\langle\overline{{\Omega^2_z({\bf
    k})}}^2\rangle\tau_{k,1}\tau_{k,2}/\langle{2\overline{\Omega_z({\bf
    k})}}^2\rangle\ll 1,
\label{weak_limit}
\end{equation} the second term is dominant. If $\Omega\tau_{k,1}$ appearing in the denominator in the second term is much larger (smaller) than 1, the SRT $\tau_{{\rm s}z}\approx\langle{\overline{\Omega_z({\bf
    k})}}^2/(2\Omega^2\tau_{k,1})\rangle^{-1}$ ($\tau_{{\rm s}z}\approx\langle{\overline{\Omega_z({\bf
    k})}}^2\tau_{k,1}/2\rangle^{-1}$) is proportional (inversely
proportional) to the momentum relaxation time. By noticing that the regime
  where Eq.~(\ref{weak_limit}) is satisfied is also in the
  strong scattering limit for the case of the weak effective Zeeman magnetic field ($\langle|\Omega_z({\bf
  k})|\rangle\tau_{k,1}\lesssim 1$), where the SRT is inversely proportional to the momentum relaxation time, therefore, the behavior under the strong effective Zeeman magnetic field is anomalous (normal) when $\Omega\tau_{k,1}\gg 1$ ($\Omega\tau_{k,1}\ll 1$) and the system is in the anomalous EY-like (normal strong scattering) regime.\cite{Zhou} 

For the transverse spin relaxation in the weak scattering limit, i.e., $\langle|\overline{{\Omega^2_z({\bf k})}}|/{(2\Omega)}\rangle\tau_{k,2}\gg 1$, the momentum scattering provides a spin
relaxation channel and the SRT is positively proportional to the momentum relaxation time. The system is hence in the normal weak scattering regime.\cite{Zhou} 

\begin{figure}[ht]
  {\includegraphics[width=8cm]{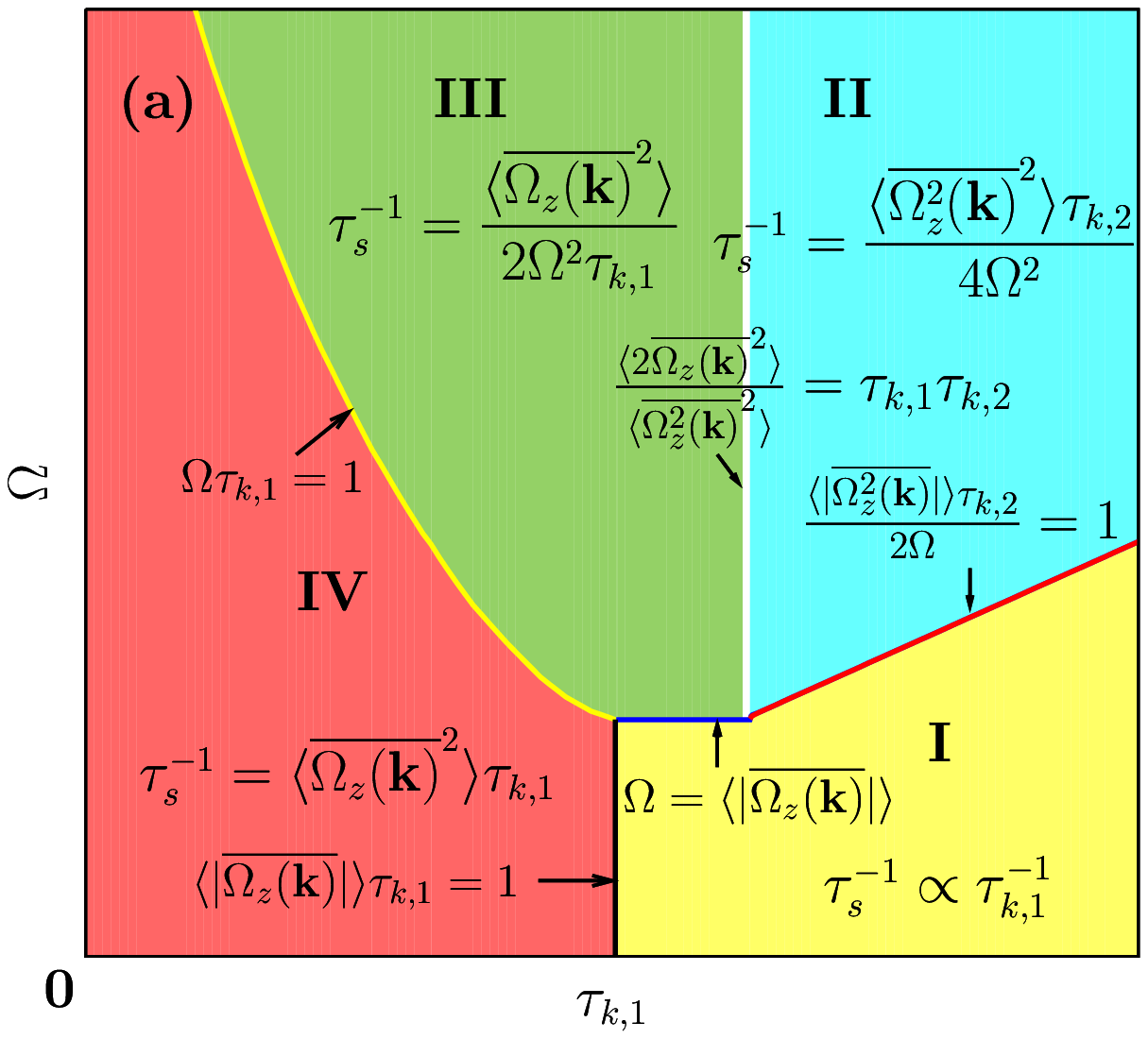}}
  {\includegraphics[width=8cm]{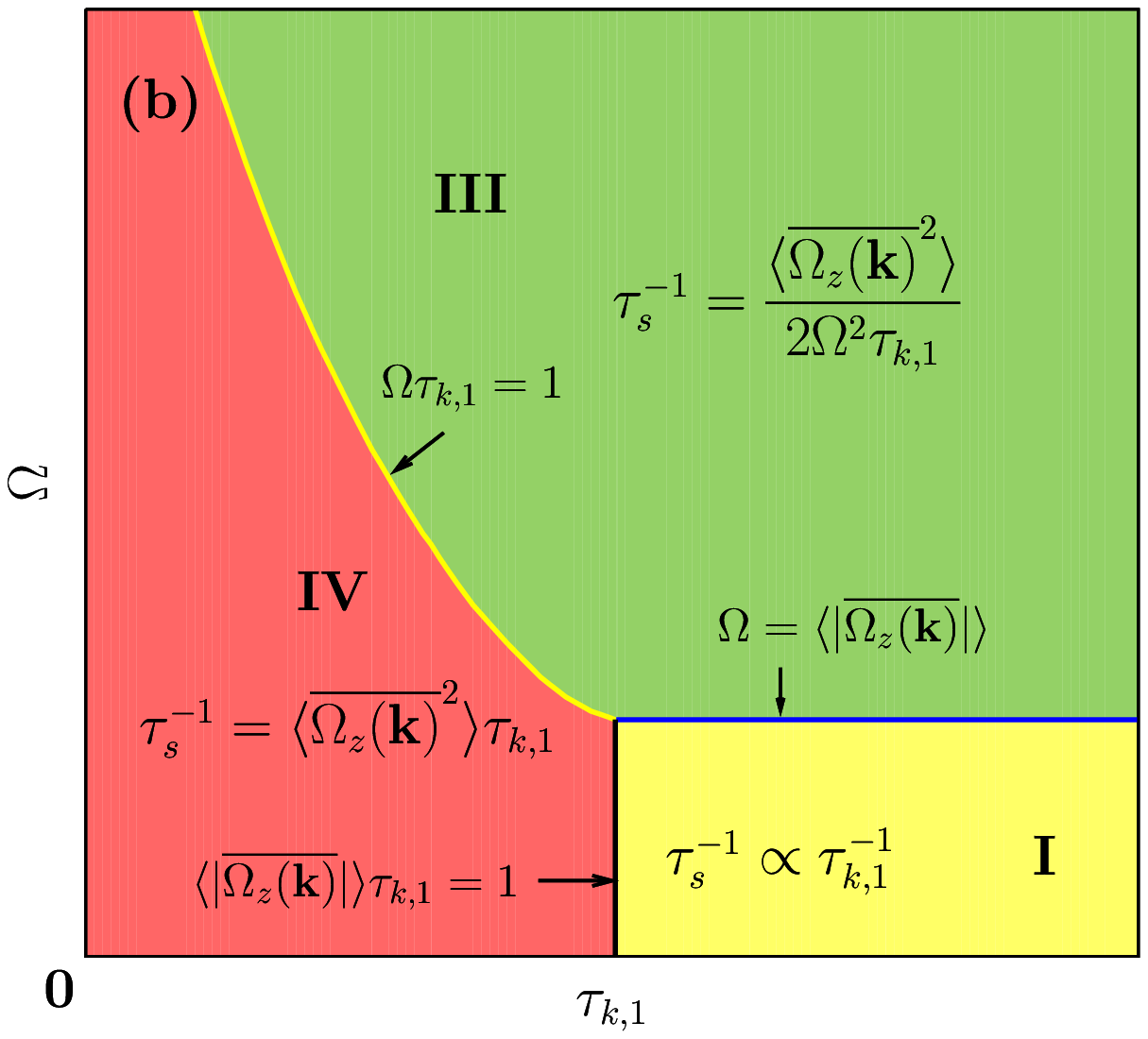}}
  \caption{(Color online) Schematic diagrams for the transverse (a) and
    longitudinal (b) spin relaxation rates
    with the effective Zeeman
magnetic field and the momentum relaxation time. The Roman numbers represent different regimes of the spin
  relaxation: I, the normal weak scattering regime; II, the anomalous DP-like regime; III,
the anomalous EY-like regime; IV, the normal strong scattering regime. The
  boundaries between the regimes I/II, II/III and III/IV are determined by Eqs.~(\ref{valley_weak}), (\ref{peak_DP}) and
(\ref{valley_EY}), respectively. The boundaries between the regimes I/II and III/IV are
determined by $\Omega=\langle|\Omega_z({\bf
  k})|\rangle$ and $\langle|\Omega_z({\bf
  k})|\rangle\tau_{k,1}= 1$, respectively.}
\label{figyw2}
\end{figure}

Therefore, with the decrease of the momentum relaxation time, the system experiences the normal weak
scattering regime ($\tau_{{\rm s}z}\propto \tau_{k,1}$), the anomalous DP-like
regime ($\tau_{{\rm s}z}^{-1}\propto \tau_{k,2}$), the anomalous EY-like regime
($\tau_{{\rm s}z}\propto \tau_{k,1}$) and the normal strong scattering regime
($\tau_{{\rm s}z}^{-1}\propto \tau_{k,1}$), successively [refer to
Fig.~\ref{figyw4}(a)], with only the weak scattering regime being in the weak
scattering limit. Accordingly, at the crossover between the normal weak
scattering and anomalous DP-like regimes, a basin may appear with the position determined by
\begin{equation} 
\langle|\overline{{\Omega^2_z({\bf k})}}/{(2\Omega)}|\rangle\tau_{k,2}\approx 1;
\label{valley_weak}
\end{equation}
 at the crossover between the anomalous DP- and anomalous EY-like regimes, a peak may arise with the position determined by 
\begin{equation}
\tau_{k,1}\tau_{k,2}=\langle2\overline{\Omega_z({\bf k})}^2\rangle/\langle\overline{\Omega_z^2({\bf k})}^2\rangle;
\label{peak_DP}
\end{equation}
and at the crossover between the anomalous EY-like and normal strong scattering regimes, a basin may arise with the position determined by 
\begin{equation}
\tau_{k,1}=\Omega^{-1}.
\label{valley_EY}
\end{equation}
The analytical results for the tansverse spin relaxation rates ($1/\tau_s$)
  with the effective Zeeman magnetic field and the momentum relaxation time in different regimes are
  summarized in the schematic diagram [Fig.~\ref{figyw2} (a)].

For the longitudinal spin relaxation, from Eq.~(\ref{x_relaxation}), when $\Omega\tau_{k,1}\gg 1$ ($\Omega\tau_{k,1}\ll 1$), the SRT $\tau_{{\rm s}x}\approx\langle{\overline{\Omega_z({\bf
    k})}}^2/(\Omega^2\tau_{k,1})\rangle^{-1}$ [$\tau_{{\rm s}x}\approx\langle{\overline{\Omega_z({\bf
    k})}}^2\tau_{k,1}\rangle^{-1}$] is proportional (inversely proportional) to
the momentum relaxation time and the system is in the anomalous EY-like (normal
strong scattering) regime [refer to Fig.~\ref{figyw4}(b)]. Moreover, there is no
normal weak scattering regime due to the efficient suppression of the strong
effective Zeeman magnetic field to the spin relaxation.\cite{Zhou} Therefore,
with the decrease of the momentum relaxation time, a basin
may appear at the crossover between the anomalous EY-like and normal strong
scattering regimes with the position determined by Eq.~(\ref{valley_EY}).
The analytical results for the longitudinal spin relaxation rates 
  with the effective Zeeman magnetic field and the momentum relaxation time in different regimes are
  summarized in the schematic diagram [Fig.~\ref{figyw2} (b)].

\subsubsection{SRT in the presence of HF effective magnetic field}
When the spin polarization is large, the HF effective magnetic field may influence the spin relaxation, as reported in semiconductors with weak Zeeman magnetic field.\cite{jianhua23,jianhua36,jianhua37,jianhua46} To analytically reveal the effect of the HF effective magnetic field on the spin relaxation with strong Zeeman magnetic field in cold atoms, we discuss the situation with $|{\bf \Omega}_{\rm HF}|\ll |\Omega|$, in which the total spin vector in Eq.~(\ref{HF}) can be substituted by ${\bf S}^{\rm T}(t)$ [${\bf S}^{\rm L}(t)$] [Eqs.~(\ref{TT}) and (\ref{LL})] for the transverse (longitudinal) spin relaxation. Accordingly, the HF effective magnetic field becomes a rotating magnetic field 
\begin{equation}
{\bf \Omega}^{\rm T}_{\rm HF}=\Omega^{\rm T}_{\rm HF}\cos(\Omega t)\hat{\bf z}-\Omega^{\rm T}_{\rm HF}\sin(\Omega t)\hat{\bf y},
\label{z_HFfield}
\end{equation}
around the effective Zeeman magnetic field when the initial spin polarization is along the ${\hat{z}}$-axis and a static magnetic field
\begin{equation}
{\bf \Omega}^{\rm L}_{\rm HF}={\Omega}^{\rm L}_{\rm HF}\hat{\bf x}
\label{x_HFfield}
\end{equation}
along the effective Zeeman magnetic field when the initial spin polarization is along the ${\hat{x}}$-axis, respectively. Here, $|\Omega_{\rm HF}|$ is the magnitude of the HF effective magnetic field.

By solving the KSBEs [Eq.~(\ref{HF_term})] with the HF effective magnetic field [Eqs.~(\ref{z_HFfield}) and (\ref{x_HFfield})] included, one obtains the transverse and longitudinal SRTs. For the transverse spin relaxation, in the strong scattering limit $\langle\Omega_{\rm eff}({\bf k})\rangle \tau_{k,2}\ll 1$ with the condition $|\Omega_{\rm HF}|\gg \langle\big|\overline{{\Omega^2_z({\bf k})}}/{(2\Omega)}\big|\rangle$, the SRT reads (the details of the derivation can be found in Appendix~\ref{AA})
\begin{equation}
\tau_{{\rm s}z}^{-1}=\left\langle\Omega^2_{\rm eff}({\bf k})\tau_{k,2}+\frac{\overline{{\Omega_z({\bf k})}}^2\tau_{k,1}}{2(1+\Omega^2\tau_{k,1}^2)}\right\rangle
\label{zz_relaxation}
\end{equation}
with
\begin{equation} 
\Omega_{\rm eff}({\bf k})\equiv{|\overline{\Omega^2_z({\bf
      k})}|}/(2\Omega\sqrt{1+\Omega^2_{\rm HF}\tau_{k,2}^2}).
\label{effective}
\end{equation}
For the longitudinal spin relaxation, the static magnetic field is added directly to the effective Zeeman magnetic field, and the longitudinal SRT becomes
\begin{eqnarray}
\tau_{{\rm s}x}^{-1}&=&\left\langle\frac{{\overline{\Omega_z({\bf
    k})}}^2\tau_{k,1}}{1+(\Omega+\Omega_{\rm HF})^2\tau_{k,1}^2}\right\rangle.
\label{xx_relaxation}
\end{eqnarray} 

It is obvious that the SRTs in Eqs.~(\ref{zz_relaxation}) and
(\ref{xx_relaxation}) can be reduced back to Eqs.~(\ref{z_relaxation}) and
(\ref{x_relaxation}) when the HF effective magnetic field $\Omega_{\rm
  HF}=0$. With the inclusion of the HF effective magnetic field, both the
transverse and longitudinal spin relaxations can be influenced. For the
transverse situation, the spin relaxation can be suppressed; for the
longitudinal situation, the spin relaxation can be either enhanced or suppressed when
the HF effective magnetic field is parallel or antiparallel to the effective Zeeman magnetic field. Therefore, when the spin polarization is large, the behavior of the spin relaxation needs to be reanalyzed.

For the transverse spin relaxation, the HF effective magnetic field in
 Eq.~(\ref{zz_relaxation}) has very different influence on the spin 
relaxation depending on whether $|\Omega_{\rm HF}|\tau_{k,2}\ll 1$ or 
$|\Omega_{\rm HF}|\tau_{k,2}\gg 1$. When $|\Omega_{\rm HF}|\tau_{k,2}\ll 1$, the spin relaxation can also be divided into similar four regimes in the absence of $|\Omega_{\rm HF}|$, but with some features modified. Specifically, the position of the peak [Eq.~(\ref{peak_DP})] at the crossover between the anomalous DP- and anomalous EY-like regimes is modified to be $\tau_{k,1}\tau_{k,2}=\langle2\overline{\Omega_z({\bf k})}^2(1+\Omega_{\rm HF}^2\tau_{k,2}^2)\rangle/\langle\overline{\Omega_z^2({\bf k})}^2\rangle$. Therefore, due to the suppression of the inhomogeneous broadening by the HF effective magnetic field, the position of the peak shows up at weaker scattering. It is also noted that, with the second term in Eq.~(\ref{zz_relaxation}) being unchanged by $|\Omega_{\rm HF}|$, the position of the basin [Eq.~(\ref{valley_EY})] at the crossover between the anomalous EY-like and normal strong scattering regimes is also unchanged. However, when $|\Omega_{\rm HF}|\tau_{k,2}\gg 1$, the HF effective magnetic field may have strong influence on the spin relaxation, with the normal weak scattering and anomalous DP-like regimes being suppressed. This is because, in this situation, as 
$\langle\Omega_{\rm eff}({\bf k})\rangle\tau_{k,2}\approx 
\langle|\overline{\Omega^2_z({\bf
    k})}|\rangle/(2\Omega\Omega_{\rm HF})$  is always much smaller
 than 1 when $|\Omega_{\rm HF}|\gg \langle\big|\overline{\Omega^2_z({\bf k})}
\big|\rangle/{(2\Omega)}$, the normal weak scattering regime is unreachable. Moreover, if $\langle\overline{\Omega^2_z({\bf
    k})}^2\rangle\tau_{k,1}\tau_{k,2}^{-1}/\langle{2\overline{\Omega_z({\bf
    k})}}^2\Omega_{\rm HF}^2\rangle\gg 1$, the first 
term in Eq.~(\ref{zz_relaxation}) is dominant and the SRT 
$\tau_{{\rm s}z}\approx\langle \overline{\Omega^2_z({\bf k})}^2/
(4\Omega^2\Omega^2_{\rm HF}\tau_{k,2})\rangle^{-1}$ is proportional to the 
momentum relaxation time, showing the EY-like behavior. Hence the 
corresponding anomalous DP-like regime in the absence of the 
HF effective magnetic field becomes the anomalous EY-like regime. 
Accordingly, with the original anomalous EY-like and normal strong scattering
 regimes being unchanged, the spin relaxation is divided into the anomalous 
EY-like and normal strong scattering regimes.       

For the longitudinal spin relaxation, the position of the basin
Eq.~(\ref{valley_EY}) at the crossover between the anomalous EY-like and normal
strong scattering regimes is modified to be $\tau_{k,1}=|\Omega+\Omega_{\rm
  HF}|^{-1}$. Therefore, with the HF effective magnetic parallel (antiparallel)
to the effective magnetic field, the basin shows up at stronger (weaker)
scattering when $|\Omega_{\rm HF}|\ll|\Omega|$.

\subsection{Numerical results}
In the previous simplified model (with the elastic scattering approximation), we are
able to calculate the spin relaxation analytically under the strong
effective Zeeman magnetic field, with the inclusion of the weak HF effective
magnetic field satisfying $|\Omega_{\rm HF}|\ll|\Omega|$. However, for genuine
system, the scattering is the inelastic interatom scattering. Moreover, the HF
effective magnetic field can be tuned to be very strong ($|\Omega_{\rm
  HF}|\gtrsim |\Omega|$). In this section, we study the SRT by solving the KSBEs
numerically. We focus on the scattering length dependence of the spin
  relaxation, with the scattering length being tuned by the Feshbach resonance.\cite{Feshbach} 

In the numerical calculation, the parameters we choose are within the experimental
feasibility by referring to
the experimental work of Wang {\it et al.}:\cite{K} The lowest two magnetic sublevels $|9/2,9/2\rangle$ and
$|9/2,7/2\rangle$ are coupled by a pair of Raman
beams with wavelength $\lambda=773$ nm and the frequency difference $\omega/(2\pi)=10.27$ MHz; The recoil momentum and energy are set to be
$k_r=k_0\sin(\theta/2)$ and $E_r=k_r^2/2m=2\pi\times 8.34\sin^2(\theta/2)$ kHz,
with $k_0=2\pi/\lambda$ and $\theta$ denoting the angle between the two Raman
beams. In our study, we choose $\sin(\theta/2)=0.125$. The Raman
detuning $\delta=\omega_z-\omega$ is set to be zero by choosing the Zeeman
shift $\omega_z/(2\pi)=10.27$ MHz. It is noted that with these parameters, the condition that the SOC energy is much
smaller than the kinetic energy is satisfied. The scattering length varies from $0.01a_0$ to $50a_0$. Here,
$a_0=169 a_{\rm B}$, with $a_{\rm B}$ being the Bohr radius, is taken from the
experimental value.\cite{K} The Fermi momentum and the temperature are set
  to be $k_{\rm F}=32 k_r$ and $T=0.2T_{\rm F}$ ($T_{\rm F}$ is the
  corresponding Fermi temperature), according to Wang {\it et al.}.\cite{K} The strengths of the effective Zeeman magnetic field $\Omega$ are taken to be $400 E_r$ and $40E_r$, corresponding to $\Omega>\langle|\Omega_z({\bf k})|\rangle$ and $\Omega<\langle|\Omega_z({\bf k})|\rangle$, respectively.

\begin{figure}[ht]
  {\includegraphics[width=7.2cm]{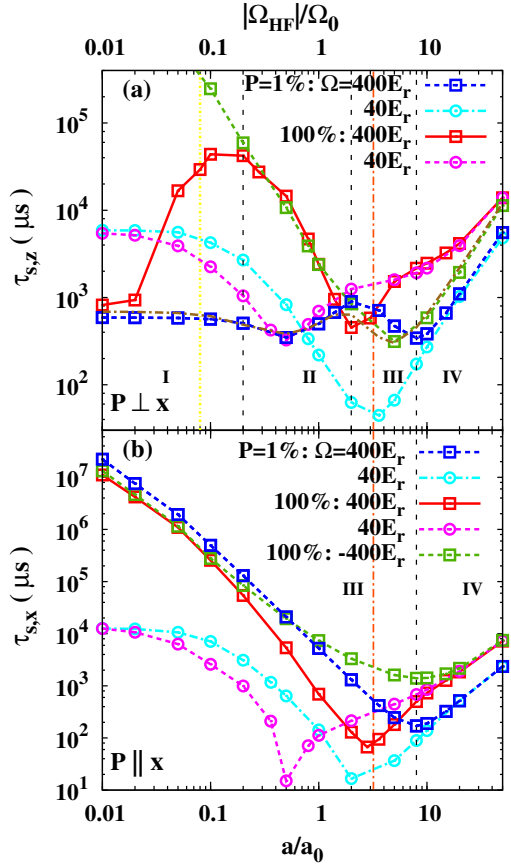}}
  \caption{(Color online) Transverse (a) and longitudinal (b)
SRTs against the scattering length under both strong and weak effective Zeeman
magnetic fields, with small and large spin polarizations $P=1\%$ and
$100\%$. The Roman numbers represent different regimes of the spin
  relaxation: I, the normal weak scattering regime; II, the anomalous DP-like regime; III,
the anomalous EY-like regime; IV, the normal strong scattering regime. The vertical
black dashed lines in (a) and (b) indicate the boundaries between different regimes for the transverse and longitudinal spin relaxations under the strong effective Zeeman magnetic
field with $P=1\%$. The vertical orange (yellow) chain lines indicate the boundary
between the weak and strong scattering regimes under the weak (strong) effective
Zeeman magnetic field with $P=1\%$ ($P=100\%$). For the transverse situation (a), the
  green dashed curve with squares represents the SRT under the {\em fixed} HF field $|\Omega|=75E_r$ and the gray chain curve shows the SRT in the absence of the HF field. For the longitudinal situation (b), $\Omega=400E_r$ ($-400E_r$) indicates the spin polarization is parallel (antiparallel) to the effective Zeeman magnetic field. We also plot the corresponding HF field $|\Omega_{\rm HF}|$ of the scattering length with $P=100\%$ (note the scale
is on the top frame of the figure with $\Omega_0=240E_r$).}
\label{figyw4}
\end{figure}

In Figs.~\ref{figyw4}(a)
and (b), the transverse and longitudinal SRTs are plotted against the
scattering length under strong and weak effective Zeeman magnetic fields,
respectively,  with both small and large spin polarizations $P=1\%$ and
$100\%$.  Below we analyze the transverse and longitudinal spin relaxations, separately. 

\subsubsection{Transverse spin relaxation}  
For the transverse spin relaxation, when the spin polarization is small ($P=1\%$), it is shown in Fig.~\ref{figyw4}(a) that the behavior of the spin relaxation under
the strong effective Zeeman magnetic field is very different
from the situation under the weak one. Under the strong effective Zeeman magnetic field, with the genuine interatom scattering, the anomalous
behaviors of the transverse spin relaxation demonstrated in the simplified
model are confirmed. Accordingly, the spin relaxation can be divided into four
regimes: I, the normal weak scattering regime; II, the anomalous DP-like regime; III,
the anomalous EY-like regime; and IV, the normal strong scattering
regime.\cite{Zhou}  Moreover, the boundaries
between the regimes I/II, II/III and III/IV can be determined by Eqs.~(\ref{valley_weak}), (\ref{peak_DP}) and
(\ref{valley_EY}) (shown by the vertical
black dashed lines), respectively. By noticing that $\tau_{k,1}$ and $\tau_{k,2}$
in these equations
are in the same order, they can be replaced by
$\tau_p^{\ast}$ limited by the interatom scattering approximately. Furthermore, by noting
that the position of the basin between III and IV is at $a=8a_0$, where
$\tau_p^{\ast}=\Omega^{-1}\propto a^{-2}$ is satisfied, we find
$\tau_p^{\ast}=(8a_0/a)^2/\Omega$. Accordingly, from Eqs.~(\ref{valley_weak})
and (\ref{peak_DP}), the boundaries
between I/II and II/III are at $a\approx 0.2a_0$ and $a\approx 2a_0$, respectively, in good agreement with the numerical calculation in Fig.~\ref{figyw4}(a). For
comparison, we also calculate the SRT under the weak effective Zeeman
  magnetic field. It can be seen from the figure that the spin relaxation can be
  divided into weak and strong scattering regimes (separated by the
  vertical orange chain line with $a\approx 3.2a_0$
  by $\langle|\Omega_z({\bf k})|\rangle\tau_p^{\ast}= 1$), which further confirms the investigation of the
  conventional spin
  relaxation in
  semiconductors under small magnetic
  field.\cite{DP,Awschalom,Zutic,fabian565,Dyakonov,Korn,wu-review}

When the spin polarization is large ($P=100\%$) and hence with large HF effective
magnetic field, it also can be seen from Fig.~\ref{figyw4}(a) that the behavior of the spin relaxation under
the strong effective Zeeman magnetic field is very different
from the situation under the weak one. In the presence of the strong effective
magnetic field, the SRT shows a peak and a basin with the
increase of the scattering length and can be divided into three
regimes rather than four. Moreover, it seems that with large spin polarization, the normal weak scattering regime
disappears and the peak (between II and III)
and the basin (between III and IV) are shifted to the weaker scattering, with the SRT showing
{\em opposite} trends in regimes I, II and III compared with the case of small spin
polarization. However, if the HF effective
magnetic field [Eq.~(\ref{HF})] is removed when $P=100\%$ (the gray chain
curve), the system also shows four regimes similar to the
case of small spin polarization, with a small modification of the positions of the peak
and the basin. This confirms that the anomalous behavior of the spin relaxation
with large spin polarization is contributed by the HF effective magnetic
field. 

One notices that the numerical results at large spin polarization differs qualitatively from the analytical analysis in Sec.~\ref{section1}. In the analytical model, with the condition $|\Omega_{\rm HF}|\gg
\langle|\overline{\Omega^2_z({\bf k})}|\rangle/{(2\Omega)}$
 satisfied, when $|\Omega_{\rm HF}|\tau_p^{\ast}\gg 1$  ($|\Omega_{\rm HF}|\tau_p^{\ast}\ll 1$), there are two (four) regimes for the spin relaxation. However, in the numerical study here, with $|\Omega_{\rm
  HF}|\tau_p^{\ast}\gg 1$ satisfied in regimes I-III (noting that with $a=0.1a_0$, $|\Omega_{\rm HF}|\tau_p^{\ast}\approx
400$ is much larger than 1), there are three instead of two regimes for the spin relaxation. This originates from the fact that in the analytical model, the HF effective magnetic field is {\em fixed} and much larger than 
$\langle|\overline{\Omega^2_z({\bf k})}|\rangle/{(2\Omega)}$. However, here the HF effective magnetic field increases {\em simultaneously} with the increase of the scattering length as $|\Omega_{\rm HF}|\propto a$ [Eq.~(\ref{HF})] and can be smaller or larger than 
$\langle|\overline{{\Omega^2_z({\bf k})}}|\rangle/{(2\Omega)}$. For comparison, the case with a {\em fixed} HF effective magnetic field $|\Omega_{\rm HF}|=75E_r$ is plotted in the figure and one finds indeed two regimes appear. It is noted that with the varying HF effective magnetic field, the system can be divided into the weak and strong scattering regimes with the boundary determined by $\langle \Omega_{\rm eff}({\bf
    k})\rangle \tau_p^{\ast}\simeq 1$ (shown by the vertical yellow chain
line with $a\approx 0.08a_0$). Below, we discuss the underlying physics in detail in the strong and weak
scattering limits, respectively.

We first focus on the strong scattering limit. In this limit, as shown in
Fig.~\ref{figyw4}(a), the SRT
first decreases and then increases with the increase of the scattering length,
showing the anomalous EY-like and normal
strong scattering behaviors, respectively. These behaviors can be
understood in the weak and strong HF effective magnetic field limits with $|\Omega_{\rm
  HF}|\ll |\Omega|$ and $|\Omega_{\rm
  HF}|\gg |\Omega|$, separately. In the weak HF effective magnetic field limit, corresponding to $a\ll 1.6a_0$
(noting that $|\Omega_{\rm HF}|\approx 400E_r$ when $a=1.6a_0$), the behavior of the spin relaxation
can be analyzed facilitated with the simplified model
[Eq.~(\ref{zz_relaxation})]. One notices that the first term in Eq.~(\ref{zz_relaxation}) is
unchanged ($\Omega_{\rm HF}^2\tau_p^{\ast}\propto a^0$) and the second term decreases with increasing the scattering
length. With $\langle\overline{{\Omega^2_z({\bf
    k})}}^2\rangle/\langle{2\overline{\Omega_z({\bf
    k})}}^2\rangle\lesssim \Omega_{\rm HF}^2$ in this regime satisfied, the second term in the
SRT is
dominant, giving the EY-like behavior in the scattering length
dependence of the spin relaxation.\cite{platform} In the strong HF effective magnetic field
limit $|\Omega_{\rm
  HF}|\gg |\Omega|$, corresponding to $a\gtrsim 10a_0$, with
$|\Omega|\tau_p^{\ast}\ll 1$, the spin relaxation returns to the
normal strong scattering regime ($\tau_s^{-1}\propto \tau_p^{\ast}$), with the SRT
increasing with the increase of the scattering length. Moreover, due to the strong HF effective magnetic field,
the inhomogeneous broadening can be effectively suppressed in this normal strong
scattering regime.\cite{jianhua23,Jianhua,jianhua36,jianhua37,wu-review} Therefore,
the SRT in this regime
is significantly enhanced compared to the situation with small spin
polarization.

We then turn to the weak scattering limit. It
is shown in Fig.~\ref{figyw4}(a) that with the increase of the scattering length, i.e., the HF
effective magnetic field, the SRT is significantly enhanced. This is because in
the presence of the strong effective Zeeman magnetic field, the inhomogeneous
broadening is markedly suppressed by the HF effective magnetic field
[Eq.~(\ref{effective})]. Consequently, a peak arises at the crossover of the weak and strong scattering limits. 

For comparison, we also calculate the SRT with the weak effective Zeeman magnetic field. The spin relaxation can also be divided into conventional weak and strong
scattering regimes, with the boundary shifted to the weaker scattering due to
the suppression of the inhomogeneous broadening by the HF effective magnetic
field.\cite{jianhua23,Jianhua,jianhua36,jianhua37,wu-review} In the strong scattering limit, the
SRT is significantly enhanced by the HF effective magnetic field, which agrees with the results in
semiconductors.\cite{jianhua23,Jianhua,jianhua36,jianhua37,wu-review} In the weak scattering limit, one observes that the
  SRT is suppressed compared to the situation with small spin polarization. This
  is because with larger spin polarization, the population of the atoms in the ${\bf k}$-space is broadened. Consequently, the inhomogeneous broadening for the spin relaxation is enhanced and the SRT is suppressed.\cite{decay}

\subsubsection{Longitudinal spin relaxation}  
For the longitudinal spin relaxation, when the spin polarization is small ($P=1\%$), it is shown in Fig.~\ref{figyw4}(b) that no matter the effective Zeeman magnetic field is strong ($|\Omega|=400E_r$) or weak ($|\Omega|=40E_r$), the SRTs always show a basin with the increase of the
scattering length.\cite{Zhou,DP,Awschalom,Zutic,fabian565,Dyakonov,Korn,wu-review} However, the underlying physics is very different. Under the strong effective Zeeman magnetic field, as revealed in the simplified model in Sec.~\ref{section1}, the spin
relaxation can be divided into the anomalous EY-like and normal strong
scattering regimes, with the boundary shown by the vertical
black dashed line at $a=8a_0$. However, under the weak effective magnetic field, the system is divided into conventional weak and strong scattering regimes (separated by the vertical purple chain line
  by $\langle|\Omega_z({\bf k})|\rangle\tau_p^{\ast}= 1$), in agreement
  with the situation in semiconductors under the weak magnetic field.\cite{Zhou,DP,Awschalom,Zutic,fabian565,Dyakonov,Korn,wu-review}

 When the spin polarization is large ($P=100\%$), the spin relaxations also present similar behaviors under the strong and weak effective Zeeman magnetic fields, showing a basin with the increase of the scattering length. In the presence of the strong effective Zeeman magnetic field, by noting that the HF effective magnetic field is antiparallel to the spin polarization [Eqs.~(\ref{HF})], when the spin
polarization is parallel (antiparallel) to the effective Zeeman magnetic
field, the total magnetic field $|\Omega+\Omega_{\rm HF}|$ along the
$\hat{x}$-axis is suppressed (enhanced) by the HF effective magnetic field. Accordingly, from Eq.~(\ref{xx_relaxation}), in which $|\Omega+\Omega_{\rm HF}|$ appears in the denominator, the SRT is suppressed (enhanced) when the spin
polarization is parallel (antiparallel) to the effective Zeeman magnetic
field, shown by the red solid (green dashed) curve with squares in the figure. 

For comparison, we also calculate the SRT in the presence of weak effective
Zeeman magnetic field. The spin relaxation can be divided into conventional weak and strong
scattering regimes, with the position of the basin shifting to the weaker
scattering due to the suppression of the inhomogeneous broadening by the HF
effective magnetic field. In the strong scattering limit, the spin relaxation
is suppressed due to the HF effective magnetic field.\cite{jianhua23,Jianhua,jianhua36,jianhua37,wu-review} In the weak scattering limit, similar to the transverse situation with weak effective Zeeman magnetic field, the SRT is suppressed due to the enhancement of the inhomogeneous broadening with large spin polarization.     

\subsubsection{Anisotropy of SRT}  
Finally, we analyze the anisotropy of the SRT under strong effective
Zeeman magnetic
field. By comparing the cases in the same conditions in Figs.~\ref{figyw4}(a)
and (b), it is shown that the SRTs for the transverse and
longitudinal spin relaxations are very different. When the spin polarization is
small, there are four regimes for the transverse spin relaxation but only two regimes
for the longitudinal situation. Specifically, at the normal strong scattering regime, the transverse SRT
is two times as large as the longitudinal one.  This anisotropy comes from
the fact that under the unique effective magnetic field
[Eq.~(\ref{field})], the inhomogeneous
broadening for the transverse and longitudinal spin relaxations are
different, with the SRTs shown in Eqs.~(\ref{z_relaxation}) and
(\ref{x_relaxation}).\cite{Zhou} When the spin polarization is large, there are three regimes for the transverse spin relaxation and again two regimes
for the longitudinal one. In this situation, the HF
effective magnetic field contributes to the anisotropy of the
transverse and longitudinal spin relaxations.

\section{Summary}
In summary, we have investigated the spin relaxation in ultracold spin-orbit coupled $^{40}$K
gas under the strong effective Zeeman magnetic field [$|\Omega|\gg
\langle|\Omega_z({\bf k})|\rangle$] both analytically and numerically. We find that
when the spin polarization is small, the SRT shows anomalous scalings with the
momentum relaxation time, with the transverse (longitudinal) spin relaxation divided into four (two) regimes in the scattering length dependence: the
normal weak scattering regime ($\tau_s\propto \tau_p^{\ast}$), the anomalous
DP-like regime ($\tau_s^{-1}\propto \tau_p^{\ast}$), the anomalous EY-like
regime ($\tau_s\propto \tau_p^{\ast}$) and the normal strong scattering regime
($\tau_s^{-1}\propto \tau_p^{\ast}$) (the anomalous EY-like regime and the
normal strong scattering regime). Specifically for the anomalous regime of the spin
relaxation, in the transverse situation, there exists a regime, i.e., the anomalous DP-like (EY-like) regime in the
conventional weak (strong) scattering limit showing anomalous spin relaxation
behavior with the SRT inversely proportional (proportional) to
the momentum relaxation time; in the longitudinal situation, the conventional
weak scattering regime is suppressed by the strong effective Zeeman magnetic
field and the system shows the EY-like behavior (i.e., the anomalous EY-like regime). 

When the spin polarization is large, a large HF effective magnetic field shows
up. We find that for the transverse spin relaxation, the HF effective
magnetic field can efficiently enhance the SRT by suppressing the inhomogeneous
broadening in both the weak and strong scattering limits. Moreover, it is noted that by
varying the scattering length, both the momentum relaxation time and the HF
effective magnetic field vary, but with $|\Omega_{\rm HF}|\tau_p^{\ast}$ fixed. Consequently, in the
scattering length dependence of the spin relaxation, when $|\Omega_{\rm HF}|\tau_p^{\ast}\gg 1$, the
anomalous DP-like regime is suppressed and the transverse SRT is
divided into three regimes: the normal weak scattering regime, the anomalous EY-like
regime and the normal strong scattering regime. Specifically for the normal weak
scattering regime, the SRT is inversely proportional to the momentum relaxation time due to the suppression of
the inhomogeneous broadening by the HF effective magnetic field. On the other
hand, for the longitudinal situation, with the inclusion of the HF effective
magnetic field, the spin relaxation is again divided into two regimes: the
anomalous EY-like regime and the normal strong scattering regime. Moreover, we
find that the spin relaxation can be either enhanced or suppressed by the HF
effective magnetic field if the spin polarization is parallel or antiparallel to the
effective Zeeman magnetic field.  

The physics of the anomalous DP spin relaxation under the
  strong effective Zeeman magnetic field revealed in this paper is rich and intriguing. Moreover, the conditions to observe
  these interesting phenomena are within the experimental
  feasibility in the
  ultracold spin-orbit coupled $^{40}$K gas.\cite{K} Till now, what predicted in this investigation has not yet been experimentally reported and we expect that our work will cause more experimental investigations.

\begin{acknowledgments}

The authors would like to thank E. Ya. Sherman for bringing this 
problem into our attention. One of the
authors (TY) would like to thank
Y. Zhou for valuable discussions. This work was supported
 by the National Natural Science Foundation of China under Grant
No. 11334014, the National Basic Research Program of China under Grant No.
2012CB922002 and the Strategic Priority Research Program 
of the Chinese Academy of Sciences under Grant
No. XDB01000000. 

\end{acknowledgments}

\begin{appendix}
\section{Analytical analysis of the transverse SRT with the HF effective
  magnetic field}
\label{AA}
We analytically derive the transverse SRT based on the KSBEs [Eq.~(\ref{HF_term})] including the HF
rotating magnetic field [Eq.~(\ref{HF})] under the
elastic scattering approximation. When the strong effective Zeeman magnetic field satisfies $|\Omega|\gg \langle|\Omega_z({\bf
  k})|\rangle$, it is convenient
to solve the KSBEs in the helix space. In the helix space, we further transform the KSBEs into the interaction picture and use the rotation wave
 and Markovian approximations.\cite{jianhua52, Haug2}

In the collinear space, the density martix can be
 expanded by the spherical harmonics 
function $Y_l^m(\theta_{\bf k},\phi_{\bf k})$
\begin{equation}
\rho_{\bf k}=\sum_{l,m}\rho_{k,l}^mY_l^m(\theta_{\bf k},\phi_{\bf k}),
\end{equation}
with $\theta_{\bf k}$ ($\phi_{\bf k}$) being the zenith (azimuth) angle between
${\bf k}$ and $\hat{x}$-axis ($\hat{y}$-axis in the $\hat{y}$-$\hat{z}$ plane).
Furthermore, by noting that the scattering term can be written as $\sum_{{l,m}}
  \rho_{k,l}^mY_l^m(\theta_{\bf k},\phi_{\bf k})/\tau_{k,l}$ [$\tau_{k,l}$ is
  defined in Eq.~(\ref{momentum})] and the ${\bf k}$-dependent magnetic field $\Omega_{z}({\bf k})$
 depends only on the zenith angle $\theta_{\bf k}$, the kinetics of the density
 martrix with $m=0$ in the KSBEs is independent on the ones with $m\ne
 0$. Therefore, we can define the quantity
\begin {equation}
\rho^{\ast}_{\bf k}=\frac{1}{2\pi}\int_0^{2\pi} d \phi_{\bf k}\,\rho_{\bf k},
\end{equation}  
which is averaged over the azimuth angle $\phi_{\bf k}$, to describe the kinetics
of the density matrix with $m=0$. Accordingly, the KSBEs
become
\begin{eqnarray}
\nonumber
&&\partial_t \rho^{\ast}_{\bf k}+{i}[\frac{\Omega_z({\bf k})}{2}\sigma_z,\rho^{\ast}_{\bf k}]+i[\frac{\Omega}{2}\sigma_x,\rho^{\ast}_{\bf
  k}]+i[\frac{{\bf \Omega}_{\rm HF}}{2}\cdot{\bgreek \sigma},\rho^{\ast}_{\bf
  k}]\\
&&\mbox{}+\sum\limits_{\bf k'}W_{\bf kk'}(\rho^{\ast}_{\bf k}-\rho^{\ast}_{\bf k'})=0.
\label{average}
\end{eqnarray}

We then transform the KSBEs [Eq.~(\ref{average})]
 from the collinear space to the helix 
one by the transformation matrix
\begin{equation}
U_{\bf k}=\left(\begin{array}{cc}
\frac{\displaystyle -\Omega}{\displaystyle \sqrt{\Omega^2+\Omega_{-}^2}} &
\frac{\displaystyle -\Omega}{\sqrt{\displaystyle \Omega^2+\Omega_{+}^2}} \\
\frac{\displaystyle \Omega_{-}}{\displaystyle
  \sqrt{\Omega^2+\Omega_{-}^2}} & \frac{\displaystyle \Omega_{+}}{\displaystyle \sqrt{\Omega^2+\Omega_{+}^2}}
\end{array}\right),
\label{UK}
\end{equation}
with $\Omega_{+}=\alpha k_x+\Omega_{\rm tot}$ and $\Omega_{-}=\alpha
k_x-\Omega_{\rm tot}$. Here, $\Omega_{\rm tot}=\sqrt{\Omega^2+\alpha^2 k_x^2}$ is half of the total
 magnetic field.
When the strong Zeeman magnetic field satisfies $|\Omega|\gg |\alpha k_x|$,
Eq.\ (\ref{UK}) can be simplified into
\begin{equation}
U_{\bf k}\approx\frac{\displaystyle 1}{\displaystyle \sqrt{2}}\left(\begin{array}{cc}
A_{\bf k}-\frac{\displaystyle \alpha k_x}{\displaystyle 2\Omega} & A_{\bf
  k}+\frac{\displaystyle \alpha k_x}{\displaystyle 2\Omega}\\
A_{\bf k}+\frac{\displaystyle \alpha k_x}{\displaystyle 2\Omega} & -A_{\bf
  k}+\frac{\displaystyle \alpha k_x}{\displaystyle 2\Omega}
\end{array}\right),
\end{equation}
with $A_{\bf k}\equiv-1+\alpha^2k_x^2/(8\Omega^2)$. 
After the transformation, the KSBEs in the helix space become
\begin{eqnarray}
\nonumber
&&\partial_t \rho_{\bf k}^h+\frac{i}{2}\sqrt{\Omega^2+\alpha^2k_x^2}[\sigma_{z'},\rho_{\bf
  k}^h]\\
\nonumber
&&\mbox{}+i \frac{\Omega_{\rm HF}}{2}\cos(\Omega t)[\sigma_{y'},\rho_{\bf k}^h]-i \frac{\Omega_{\rm HF}}{2}\sin(\Omega t)[\sigma_{x'},\rho_{\bf k}^h]\\
\nonumber
&&\mbox{}+\sum\limits_{\bf k'}W_{\bf kk'}(\rho_{\bf k}^h-\rho_{\bf
  k'}^h)\\
\nonumber
&&\mbox{}+\sum\limits_{\bf k'}W_{\bf kk'}\frac{\alpha^2(k_x'-k_x)^2}{4\Omega^2}(\rho_{\bf k'}^h-\sigma_{y'}\rho_{\bf
  k'}^h\sigma_{y'})\\
&&\mbox{}+\sum\limits_{\bf k'}W_{\bf kk'}\frac{i\alpha}{2\Omega}(k_x-k_x')[\sigma_{y'},\rho_{\bf k'}^h]=0,
\label{helix}
\end{eqnarray}
with the density matrix in the helix space with $m=0$ being
$\rho_{\bf k}^h=U_{\bf k}^{\dagger}\rho^{\ast}_{\bf
  k}U_{\bf k}$.

Equations (\ref{helix}) describe the kinetics of the density matrices in the
 helix space under the magnetic field $\Omega_{\rm tot}/2$ together
with the rotating HF
 effective magnetic field. Moreover, in the helix space, 
the scattering terms include
not only the spin-conserving part [the fifth and sixth terms in the 
left hand side (LHS) of Eq.~(\ref{helix})], 
but also the effective spin-flip part [the last term 
in the LHS of Eq.~(\ref{helix})].   
 
We then transform the KSBEs into the interaction picture with the density matrix
\begin{equation}
\tilde{\rho}_{\bf k}=e^{i\Omega_{\rm HF}\sigma_{y'} t/2}e^{i\Omega\sigma_{z'} t/2}\rho_{\bf
  k}^he^{-i\Omega\sigma_{z'} t/2}e^{-i\Omega_{\rm HF}\sigma_{y'} t/2}.
\end{equation}
Here, the density matrix 
can be expanded by the Legendre function
\begin{equation}
\tilde{\rho}_{\bf k}=\sum_{l} C_l\tilde{\rho}_{k,l} P_l(\cos \theta_{\bf k}),
\end{equation}
with $C_l=\sqrt{(2l+1)/(4\pi)}$. In the derivation below, the scattering
  matrix element $|U_{\bf k-k'}|$ in $W_{\bf kk'}$ is taken to be constant by considering the property of the interatom scattering potential [Eq.~(\ref{interaction})].
 By defining the spin vector $\tilde{{\bf S}}_{{\bf
    k}}^l={\rm Tr}[\tilde{\rho}_{k,l}\bgreek{\sigma}]$,
taking the approximation 
 $\sqrt{\Omega^2+\alpha^2k_x^2}\approx \Omega+{\alpha^2k_x^2}
/(2\Omega)$ and applying the rotation wave approximation 
($|\Omega|\gg |\alpha k_x|$), one obtains
\begin{eqnarray}
\nonumber
&&\frac{\partial \tilde{{\bf S}}_k^{l}}{\partial t}+\frac{\Omega_{\rm so}^2({\bf k})}{2\Omega}U_1(t)\Big[\frac{4l^3+6l^2-1}{(2l-1)(2l+1)(2l+3)}\tilde{{\bf S}}_k^{l}\\
\nonumber
&&\mbox{}+\sqrt{\frac{2l-3}{2l+1}}\frac{l(l-1)}{(2l-3)(2l-1)}\tilde{{\bf S}}_k^{l-2}+\sqrt{\frac{2l+5}{2l+1}}\\
\nonumber
&&\mbox{}\times\frac{(l+1)(l+2)}{(2l+3)(2l+5)}\tilde{{\bf S}}_k^{l+2}\Big]
+U_2(t)\tilde{{\bf S}}_k^{l}\\
&&\mbox{}+\frac{\sqrt{3}\Omega_{\rm so}({\bf k})}{3\Omega\tau_{k,1}}U_3(t)\big(\tilde{{\bf
  S}}_k^{0}\delta_{l1}-\tilde{{\bf S}}_k^{1}\delta_{l0}\big)
=0
\label{matrixequation},
\end{eqnarray}
with $\Omega_{\rm so}({\bf k})\equiv\alpha k$ and $\delta_{ij}$ being the Kronecker symbol. Here, the matrices
$U_1(t)$ to $U_3(t)$ are defined by
\begin{widetext}

\begin{equation}
U_1(t)=\left(\begin{array}{ccc}
0 & \cos(\Omega_{\rm HF}t) & 0\\
-\cos(\Omega_{\rm HF}t)& 0 & -\sin(\Omega_{\rm HF}t)\\
0 & \sin(\Omega_{\rm HF}t)  & 0
\end{array}\right),
\end{equation}
\begin{equation}
U_2(t)=\left(\begin{array}{ccc}
\frac{\displaystyle 1}{\displaystyle \tau_{k,l}}+\frac{\displaystyle \Omega_{\rm
    so}^2({\bf k})}{\displaystyle 4\Omega^2\tau_{k,1}}\delta_{l0} & 0 & 0\\
0 & \frac{1}{\displaystyle \tau_{k,l}}+\frac{\displaystyle \Omega_{\rm
    so}^2({\bf k})}{\displaystyle 6\Omega^2\tau_{k,1}}
\delta_{l0} & 0\\
0 & 0 &  \frac{\displaystyle 1}{\displaystyle \tau_{k,l}}+\frac{\displaystyle
  \Omega_{\rm so}^2({\bf k})}{\displaystyle 4\Omega^2\tau_{k,1}}
\delta_{l0}
\end{array}\right),
\end{equation}
and
\begin{equation}
U_3(t)=\left(\begin{array}{ccc}
0 & -\sin(\Omega t)\sin(\Omega_{\rm HF}t) & -\cos(\Omega t) \\
\sin(\Omega t)\sin(\Omega_{\rm HF}t)& 0 & -\sin(\Omega t)\cos(\Omega_{\rm HF}t)\\
\cos(\Omega t) & \sin(\Omega t)\cos(\Omega_{\rm HF}t) & 0
\end{array}\right).
\end{equation}

\end{widetext}

In Eq.~(\ref{matrixequation}), when $\Omega_{\rm
    so}^2({\bf k})\tau_{k,2}/(2\Omega)\ll 1$, we retain terms with $l\le 2$. Specifically, in Eq.~(\ref{matrixequation}), we have retained terms within the lowest two orders in the last two terms.
By further applying the Markovian approximation, the tansverse spin vector in the interaction picture reads
\begin{equation}
\tilde{{\bf S}}_k^{0}(t)\approx p({\bf k})\exp\big[{-t/\tau_{sz}({\bf k})}\big], 
\end{equation}
with $p({\bf k})=\mbox{Tr}[\rho_{\bf
    k}(0)\cdot {\bgreek \sigma}]/n_a$ being the initial spin vector for the atom with momentum ${\bf k}$. The transverse SRT $\tau_{sz}({\bf k})$ is therefore 
\begin{eqnarray}
\nonumber
\tau_{sz}^{-1}({\bf k})&=&\frac{\Omega^4_{\rm so}({\bf
    k})\tau_{k,2}}{45\Omega^2(1+\Omega^2\tau_{k,2}^2)}+\frac{\Omega^2_{\rm so}({\bf k})}{12\Omega^2}\Big[\frac{(\Omega+\Omega_{\rm HF})^2\tau_{k,1}}{1+(\Omega+\Omega_{\rm
    HF})^2\tau_{k,1}^2}\\
&+&\frac{(\Omega-\Omega_{\rm
    HF})^2\tau_{k,1}}{1+(\Omega-\Omega_{\rm
    HF})^2\tau_{k,1}^2}\Big].
\label{tausz}
\end{eqnarray}
By considering $|\Omega_{\rm HF}|\ll |\Omega|$ in our derivation, the above equation
 is further simplified to be
\begin{equation}
\tau_{sz}^{-1}({\bf k})=\frac{\Omega^4_{\rm so}({\bf
    k})\tau_{k,2}}{45\Omega^2(1+\Omega_{\rm HF}^2\tau_{k,2}^2)}+\frac{\Omega^2_{\rm so}({\bf k})\tau_{k,1}}{6(1+\Omega^2\tau_{k,1}^2)}.
\end{equation}

\end{appendix}


\begin{thebibliography}{0}
\bibitem{platform}  If the first term in Eq.~(\ref{zz_relaxation}) is
dominant, the SRT shows a platform in the scattering length dependence.
\bibitem{decay}  Under the unique effective magnetic field ${\bf{\Omega(k)}}=({\Omega},0,\alpha k_x)$ with $|\Omega|\ll \langle|\Omega_z(\bf k)|\rangle$, the spin polarization, i.e., the HF effective magnetic field is roughly along the $\hat{z}$-axis. Therefore, when  $|\Omega_{\rm HF}|\ll\langle|\Omega_{z}({\bf k})|\rangle$, the HF effective magnetic field has little influence on the spin relaxation. Indeed, if we remove the HF magnetic field in the computation at this regime, the SRT changes little.  By solving the KSBEs [Eq.~(\ref{HF_term})] with the HF effective magnetic field and the scattering term neglected, one finds the spin vector along the $\hat{z}$-direction 
\begin{equation}
\nonumber
S_z(t)=\sum_{\bf k}\frac{p({\bf k})}{\alpha^2k_x^2+\Omega^2}\big(\Omega^2\cos(\sqrt{\alpha^2k_x^2+\Omega^2}t)+\alpha^2k_x^2\big),
\end{equation}
with $p({\bf k})=\mbox{Tr}[\rho_{\bf
    k}(0)\cdot {\bgreek \sigma}]/n_a$ being the initial spin vector for the atom with momentum ${\bf k}$. Accordingly, when the population for the atoms in ${\bf k}$-space is broadened, the inhomogeneous broadening $\langle\alpha^2k_x^2\rangle$ is enhanced.
\end{thebibliography}
\end{document}